\documentclass[aps,prd,twocolumn,nofootinbib,superscriptaddress,preprintnumbers]{revtex4}
\usepackage[hypertex]{hyperref}
\usepackage{graphicx}% Include figure files
\usepackage{mciteplus}
\usepackage{afterpage}

\begin{document}

\pagestyle{plain}

\preprint{}

\title{Signals of Inert Doublet Dark Matter in Neutrino Telescopes}
\author{Prateek Agrawal}
\affiliation{Department of Physics, University of Maryland, College Park, 
MD 20742}
\author{Ethan M. Dolle}
\affiliation{Department of Physics, University of Arizona, Tucson, AZ 85721}
\author{Christopher A. Krenke}
\affiliation{Department of Physics, University of Maryland, College Park, 
MD 20742}
\affiliation{Department of Physics, University of Arizona, Tucson, AZ 85721}

\begin{abstract}

One of the simplest extensions of the Standard Model that explains the observed abundance of dark matter is 
the inert doublet model. In this theory a discrete symmetry ensures that the neutral component of 
an additional electroweak doublet scalar is stable, and constitutes a dark matter candidate. As massive 
bodies such as the Sun and Earth move through the dark matter halo, dark matter particles 
can become gravitationally trapped in their cores.  Annihilations of these particles result in 
neutrinos, which can potentially be observed with neutrino telescopes.  We 
calculate the neutrino detection rate at these experiments from inert doublet dark matter annihilations in the cores
of the Sun and the Earth.  

\end{abstract}

 \maketitle

\section{Introduction}
One of the most pressing questions currently in physics is the nature
of dark matter.  It is known that most of the matter in the universe
is comprised of non-luminous (dark), neutral matter.  Moreover, it is
also known that none of the standard model particles can be the dark
matter.  Hence, to explain dark matter, we must introduce new physics
beyond the standard model. The most appealing scenario from a
cosmological standpoint is that of a weakly interacting massive
particle (WIMP).  The WIMP is theoretically attractive because any
weakly interacting particle with a mass in the TeV scale will give
approximately the correct relic abundance of dark matter.  Many
extensions of the standard model contain such dark matter candidates,
for example, supersymmetry \cite{Ellis84, *Griest88}, extra dimensions
\cite{Kolb84,Dienes99,*Servant03} and little Higgs
\cite{Birkedal-Han04}.

Dark matter is distributed across the galaxy in the form of a halo.
As the dark matter particles interact with nuclei in large objects
such as the Sun or Earth, a certain fraction of the scattered dark
matter particles will not have enough kinetic energy to escape the
gravitational pull of the object.  Thus, eventually a build up of dark
matter will occur in the cores of massive objects.  The dark matter
particles will then annihilate, producing standard model particles,
which eventually decay to neutrinos.  If the neutrinos originate from
the cascade of a TeV scale object such as a WIMP, they will typically
have an energy of a few hundred GeV. This energy range is detectable
in neutrino telescopes such as IceCube. This idea has been
investigated before in the context of supersymmetric dark matter
\cite{Silk85} and Kaluza-Klein dark matter candidates
\cite{Cheng02,*Hooper03}.

In this paper, we consider the possible signal in neutrino telescopes
from dark matter annihilations in the Sun and Earth for the inert
doublet model (IDM). In this model, the Higgs sector of the standard
model (SM) is extended to include an additional $SU(2)$ doublet that
does not acquire a vacuum expectation value (vev).  If an additional
$Z_2$ symmetry is imposed, the neutral particle in the additional
doublet becomes stable and thus, a dark matter candidate. The possible
neutrino signal from the inert doublet model is especially interesting
because this type of dark matter arises in extensions of the SM
motivated by the little hierarchy problem, such as the Left-Right Twin
Higgs model \cite{Chacko06,*Goh07}.

The format of the paper is as follows:  In Section II we discuss the
basics of the IDM and list some constraints.  In section III we
discuss the dynamics of capture of dark matter particles by massive
objects and calculate the capture and annihilation rates for the Sun
and Earth for the inert doublet model.  In Section IV, we discuss the
determination of the detector rate.  In section V we present our
results, and in section VI we conclude.

\section{The Inert Doublet Model} The inert doublet model consists of
the standard model with an additional electroweak doublet scalar,
$H_2$, with the same quantum numbers as the SM Higgs.  This doublet is
odd under a $Z_2$ symmetry, while all other fields are even, and does
not acquire a vev.  Hence, $H_2$ couples only to SM gauge bosons, the
SM Higgs and itself.  Since $H_2$ is odd under the $Z_2$ symmetry,
this ensures that the lightest component of $H_2$ is stable.  Thus, if
the neutral scalar is lighter than the charged scalar in $H_2$, it is
a good dark matter candidate \cite{Chacko06,Ma06}.

The general pattern of symmetry breaking in a two Higgs doublet model
was first investigated by Deshpande and Ma \cite{Deshpande78}.
Recently, Barbieri et al. have shown that in the case of an unbroken
$Z_2$ and appropriately chosen splittings the model passes electroweak
precision tests with a heavy Higgs, thus solving the little hierarchy
problem \cite{Barbieri06}. They refer to this case as the inert
doublet model since the new Higgs doublet cannot couple to fermions.
The IDM has been shown to give the correct relic abundance for dark
matter \cite{Dolle08}. The direct detection rates of dark matter in
this model were examined by \cite{Majumdar07}. It has also been
analysed for indirect signals of dark matter via photons
\cite{Honorez07} and monochromatic photon production at the Galactic
center \cite{Gustafsson07}. The LEP II analysis for neutralino pair
production has recently been translated into constraints on the IDM
\cite{Lundstrom08}.

\subsection{Parametrization of the Potential}

The most general potential consistent with the $Z_2$ symmetry is
\begin{eqnarray} V&=&\mu_1^2|H_1|^2+\mu_2^2|H_2|^2+\lambda_1|H_1|^4
  \nonumber \\ &+&\lambda_2|H_2|^4+\lambda_3|H_1|^2|H_2|^2 \nonumber
  \\ &+&\lambda_4|H_1^{\dagger}H_2|^2+\frac{\lambda_5}{2}\left \{
  (H_1^{\dagger}H_2)^2+h.c. \right \}.  \end{eqnarray}

Expanding the potential above in unitary gauge,

\begin{eqnarray} H_1 &=& \left( \begin{array}{c}
  0 \\ (v+h)/\sqrt{2}\\ \end{array} \right) \\ H_2 &=& \left(%
    \begin{array}{c} H^+ \\ (S+iA)/\sqrt{2}\\ \end{array} \right),
    \end{eqnarray}

we obtain, 

\begin{eqnarray} m_h^2 &=& 2v^2\lambda_1=-2\mu_1^2 \nonumber \\
  m_{S}^2 &=& \mu_2^2+\frac{1}{2}(\lambda_3+\lambda_4+\lambda_5)v^2
  \nonumber \\ m_{A}^2 &=&
  \mu_2^2+\frac{1}{2}(\lambda_3+\lambda_4-\lambda_5)v^2 \nonumber \\
  m_{H^{\pm}}^2 &=& \mu_2^2+\frac{1}{2}\lambda_3v^2.  \end{eqnarray}

This leaves us with seven independent real parameters: $\mu_1$,
$\mu_2$, $\lambda_1$, $\lambda_2$, $\lambda_3$ , $\lambda_4$, and
$\lambda_5$.  Fixing the Z mass fixes $v$, while fixing the SM Higgs
mass fixes $\mu_1$ and $\lambda_1$.  Following \cite{Barbieri06} we
define $\lambda_L=(\lambda_3+ \lambda_4+\lambda_5)$.  Fixing the
scalar mass and the mass splittings between the scalar, pseudoscalar,
and charged particles as $m_{S}$, $\delta_1=m_{H^{\pm}}-m_{S}$ and
$\delta_2=m_{A}-m_{S}$, fixes $\mu_2$, $\lambda_3$, and $\lambda_5$.
This leaves us with the new parameter set: $m_Z$, $m_h$, $\lambda_2$,
$\lambda_L$, $\delta_1$, $\delta_2$, and $m_S$.  The $Z$ mass is fixed
by LEP, while we take $\lambda_2=0.1$ \cite{Honorez07}. This parameter
does not affect the neutrino flux rate calculation directly.

\subsection{Constraints} There are a number of constraints that serve
to limit our parameter space.  We discuss each of them in detail
below.

\subsubsection{From $\Omega h^2$} Recent measurements from WMAP have
led to a precise determination of the amount of dark matter in the
Universe: $\Omega_{DM} h^2=0.112 \pm 0.009$ \cite{Spergel07}.  The
program micrOMEGAs \cite{Micromegas} uses a calcHEP \cite{Calchep}
model file to solve the Boltzman equation numerically to find the
relic density.  We used this to exclude a large portion of the
parameter space.  Two regions are consistent with WMAP at the $3
\sigma$ level: a low mass region, where $m_{S}<$ 100 GeV, and a high
mass region, where 500 GeV $<m_{H_0}<$ 2 TeV.

\subsubsection{From direct detection bounds} The direct detection
bounds put a constraint on the value of $\lambda_5$. As
$\lambda_5\rightarrow0$, the neutral scalar $S$ and the pseudo-scalar
$A$ become degenerate. In this case, these particles can scatter off
matter via a $Z$ exchange, giving cross-sections eight to nine order
of magnitudes larger than present direct detection bounds
\cite{akerib06}. This constraint can be avoided if the mass splitting
between $S$ and $A$ is higher than the kinetic energy of dark-matter
in the halo, or $\delta_2\sim$ a few hundred MeV.

\subsubsection{From $\Gamma_Z$} Data from LEP has put tight
constraints on the width of the Z boson, such that new decay channels
are very restricted. This puts lower bounds on the scalar mass and the
mass splittings.  \begin{eqnarray} m_{S}+m_{A}>m_Z &\Rightarrow&
  m_{S}>\frac{m_Z-\delta_2}{2} \nonumber\\ 2m_{H^{\pm}}>m_Z
  &\Rightarrow& m_{S}>\frac{m_Z-2\delta_1}{2} \nonumber \end{eqnarray}

\subsubsection{From model stability} In order to ensure stability of
the model, we require $\lambda_L,
\lambda_3>-2\sqrt{\lambda_1\lambda_2}$.  Since we consider the scalar,
$S$, to be the dark matter candidate, $\lambda_L<\lambda_3$, and the
first condition is sufficient.  Substituting the SM Higgs mass for
$\lambda_1$ yields the condition:
$\lambda_L>-\frac{2m_h}{v}\sqrt{\frac{\lambda_2}{2}}$.  For a modest
choice of $\lambda_2$ and $m_h$, this excludes almost all negative
$\lambda_L$ couplings.

\section{Dark Matter Capture and Annihilation} Dark matter particles
(DMPs) accumulate in the massive bodies from the Galactic halo, and
are depleted by annihilations. Given a long enough time, this process
can come into equilibrium. The differential equation governing the
number of DMPs in the Sun (or Earth) is 

\begin{equation} \dot{N}=C-C_{A}N^{2}\label{eq:diffeq}\end{equation}

Here $C$ is the capture rate from the halo, and
$C_{A}=\langle\sigma_{A}v\rangle V_{2}/V_{1}^{2}$.
$\langle\sigma_{A}v\rangle$ is the total cross-section times the
relative velocity in the limit $v\rightarrow0$. $V_{j}$ are the
effective volumes for the Sun and Earth, given by 

\begin{equation}
  V_{j}=(3m_{pl}^{2}T/(2jm_{S}\rho))^{3/2}\end{equation}

and $m_{pl}$ is the Planck mass, $T$ is the temperature and $\rho$ the
core density of the massive body. Equation (\ref{eq:diffeq}) can be
solved to obtain the annihilation rate of DMPs

\begin{equation}
  \Gamma_{A}=\frac{1}{2}C\tanh^{2}(t/\tau)\end{equation}

If the time scale for equilibriation ($\tau=\sqrt{CC_{A}}$) is much
smaller than the age of the solar system ($t$), then $\Gamma_{A}$ is
simply $\frac{1}{2}C$. Since this is the scenario of a maximal signal,
this indicates that the capture rate is the dominant factor in
determining the magnitude of the signal. 

\subsection{Capture Rate} The capture rate depends on the DMP
scattering cross-section from the nuclei. The calculation of
cross-section proceeds in three steps.  First we calculate the
partonic cross-section with quarks and gluons.  We next translate this
into the interaction with nucleons, by taking the quark/gluon matrix
elements in the nucleonic state. Finally, we evaluate the nucleon
operator matrix elements in the nucleus \cite{SHIFMAN78}. Since these
particles are non-relativistic, the cross-section calculation is
greatly simplified. The DMP undergoes scalar interactions which
coherently add in the nucleus, and hence it couples to the mass of the
nucleus. The elastic cross-section at zero momentum transfer is
\cite{Barbieri06}

\begin{equation}
  \sigma_{0}^{i}=\frac{m_{S}^{2}m_{N_{i}}^{2}}{4\pi(m_{S}+m_{N_{i}})^{2}}
  \left(\frac{\lambda_{L}}{m_{S}m_{h}^{2}}\right)^{2}f^{2}m_{N_{i}}^{2}\end{equation}

Here, $m_N$ is the mass of the nucleus and $f\sim0.3$ is the nucleonic
matrix element \cite{Ellis00}, defined by \begin{equation} \langle
  N|\sum m_{q}q\bar{q}|N\rangle=fm_{N}\langle N|N\rangle.
\end{equation}

At a finite momentum transfer, the particle doesn't see the entire
nucleus, and hence the cross-section is suppressed by a form-factor
($F_{i}(m_{S})$). Other factors that influence the capture rate are
the elemental abundance $f_{i}$, distribution $\phi_{i}$ and the
kinematic suppression, $S(m_{s}/m_{N_{i}})$ \cite{Gould1987}. The
capture rate is given by

\begin{eqnarray}
  C&=&c\frac{\rho_{0.3}}{(m_{S}/\text{GeV})\bar{v}_{270}}\sum_{i}F_{i}(m_{S})
  \left(\frac{\sigma_{0}^{i}}{10^{-4} \text{ pb}}\right) \nonumber \\
  &\times&
  f_{i}\phi_{i}\frac{S(m_{S}/m_{N_{i}})}{(m_{N_{i}}/\text{GeV})}
\end{eqnarray} where $c=4.8\times10^{24}$ s$^{-1}$ for the Sun and
$c=4.8\times10^{15}$ s$^{-1}$ for Earth, $\rho_{0.3}$ is the local
halo mass density in units of $0.3\ \text{GeV cm}^{-3}$, and
$\bar{v}_{270}$ is the dark-matter velocity dispersion in terms of
$270\ \text{km s}^{-1}$. The quantities $f_{i}$ and $\phi_{i}$ for the
Sun and Earth are given in \cite{jungman-1996-267} The sum is over all
elements in the Sun or Earth.  The kinematic suppression factor can be
parametrized as following

\begin{eqnarray}
  S(x)&=&[A^{\frac{3}{2}}/(1+A^{\frac{3}{2}})]^{\frac{2}{3}}\label{eq:kinsup1}\\
  A&=&\frac{3}{2}\frac{x}{(x-1)^{2}}\left(\frac{\langle
  v_{esc}\rangle^{2}}{\bar{v}^{2}}\right)\label{eq:kinsup2}
\end{eqnarray}

The mean escape velocity, $\langle v_{esc}\rangle$ is obtained by
numerical fitting to the exact kinematic suppression. For the Sun,
$\langle v_{esc}\rangle=1156\text{ km s}^{-1}$, and for the Earth
$\langle v_{esc}\rangle=13.2\text{ km s}^{-1}$.  This factor arises
from the fact that in order to be captured, the scalar has to scatter
off a nucleus with a velocity less than the escape velocity in that
object. If the mass of the nucleus is equal to the mass of the scalar,
then it can potentially lose all its kinetic energy in an elastic
collision. It can be seen from the parametrization above that the
capture is suppressed more if the mass of the dark matter particles is
very different from the mass of the nucleus.  The form factor
suppression can be modeled numerically for simplicity.  For the Earth,
the only appreciable effect is the suppression from iron, so
$F_{i}(m_{S})\simeq1$ for all other elements. For iron,

\begin{equation}
  F_{Fe}\simeq1-0.26\left(\frac{A}{A+1}\right)\end{equation}

For the Sun, the fitting is slightly more complicated. The suppression
is given by

\begin{equation}
  F_{i}(m_{S})=F_{i}^{inf}+(1-F_{i}^{inf})\exp\left[-\left(\frac{\log(m_{s})}{\log(m_{c}^{i})}
  \right)^{\alpha_{i}}\right]\end{equation}

The quantities $F_i^{inf}$, $m_c^i$ and $\alpha_i$ are numerical
fitting parameters taken from \cite{jungman-1996-267}.

\begin{figure*}[htbp] \centering
  \includegraphics[width=0.45\textwidth]{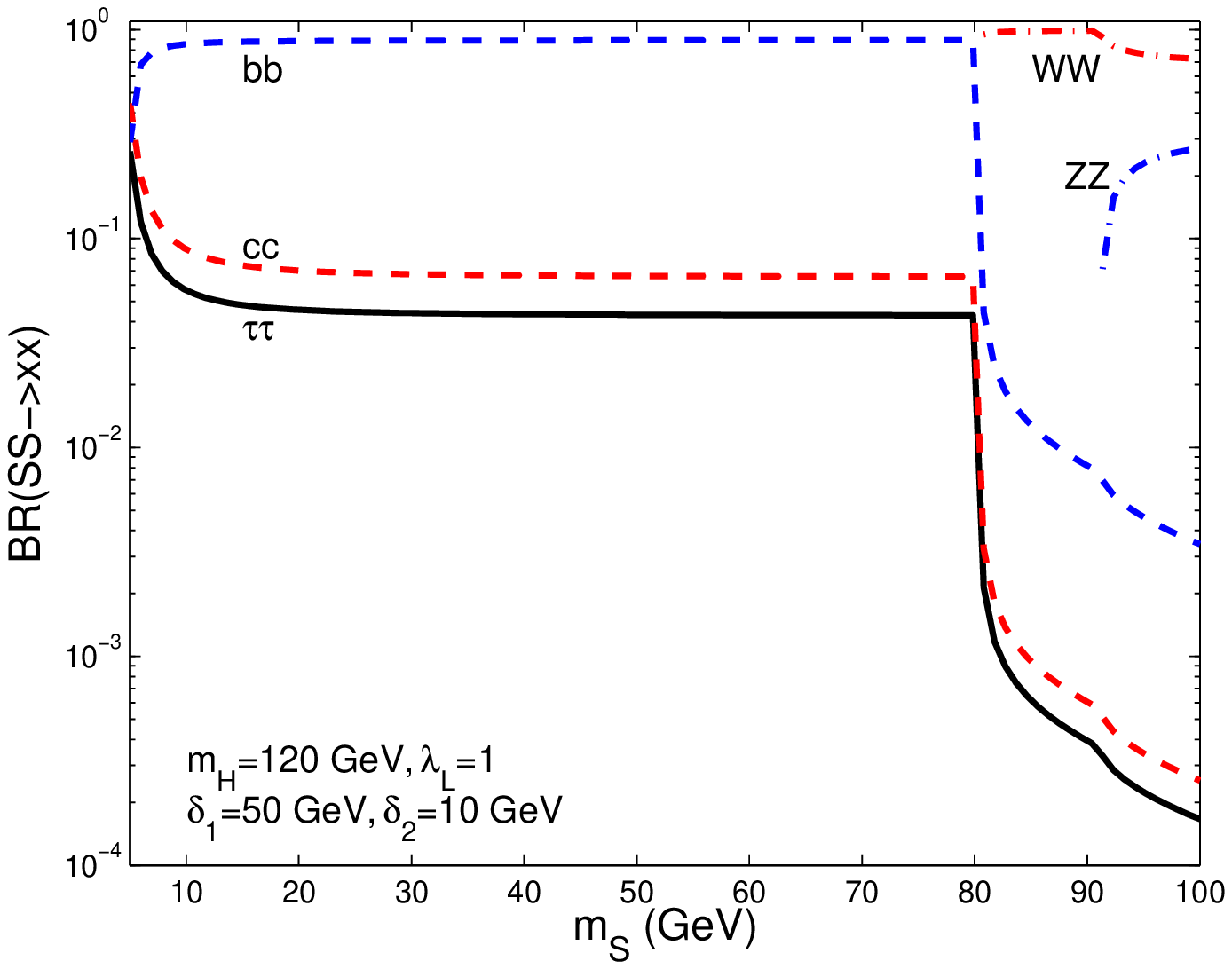}
  \includegraphics[width=0.45\textwidth]{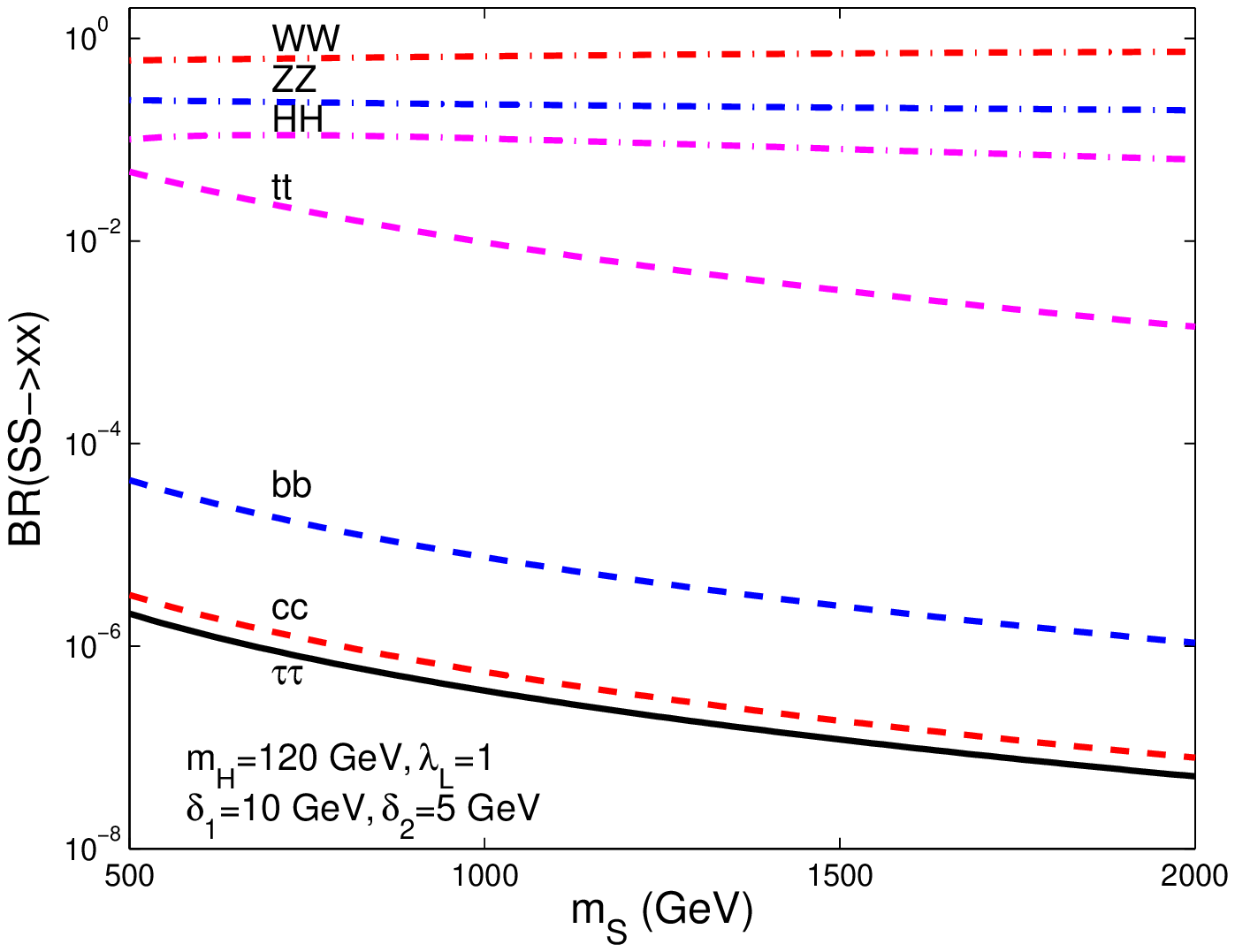}
  \caption{Branching Fractions of S into SM particles for low mass
  region (left) high mass region (right).}\label{fig:brfr}

\includegraphics[width=0.30\textwidth]{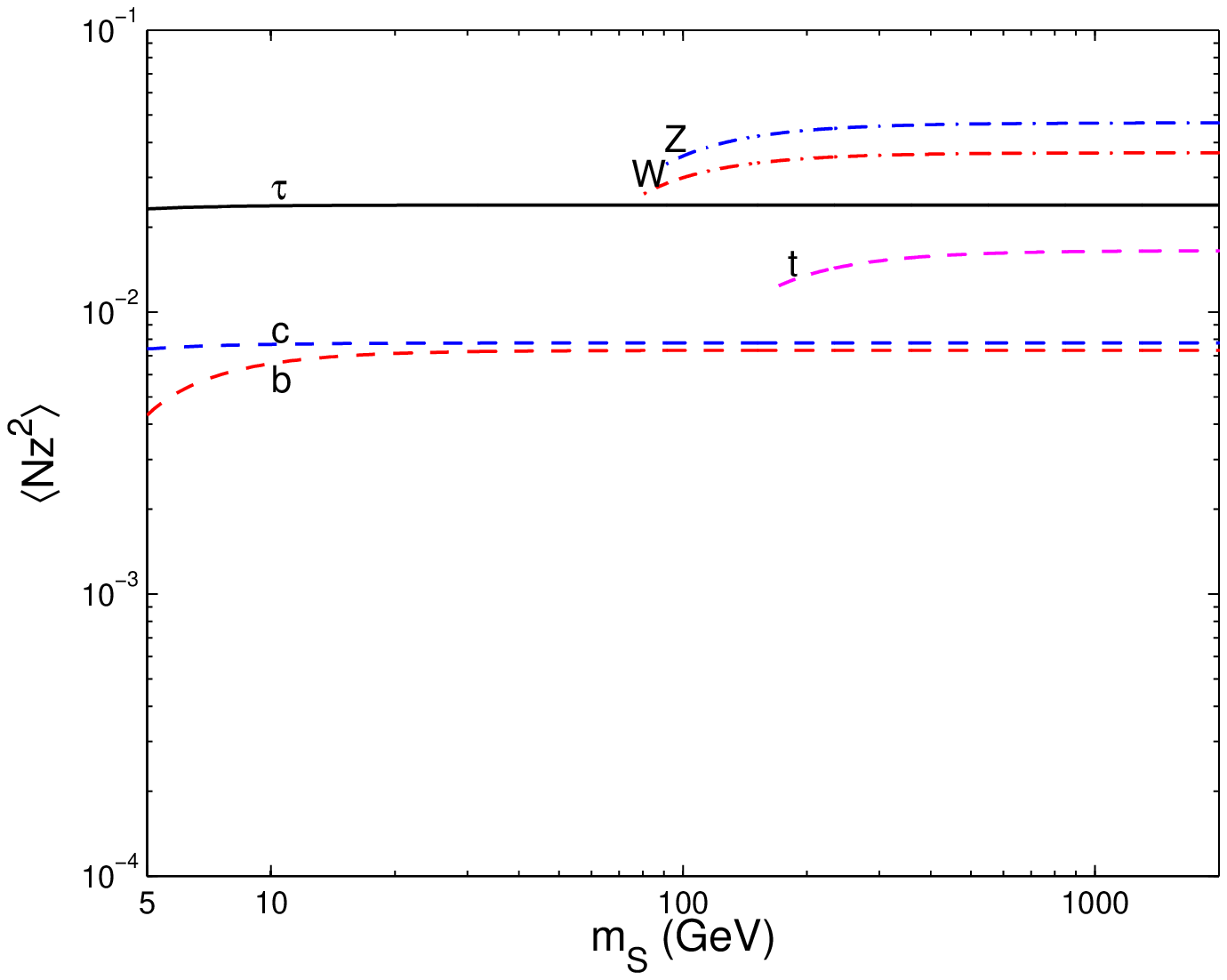}
\includegraphics[width=0.30\textwidth]{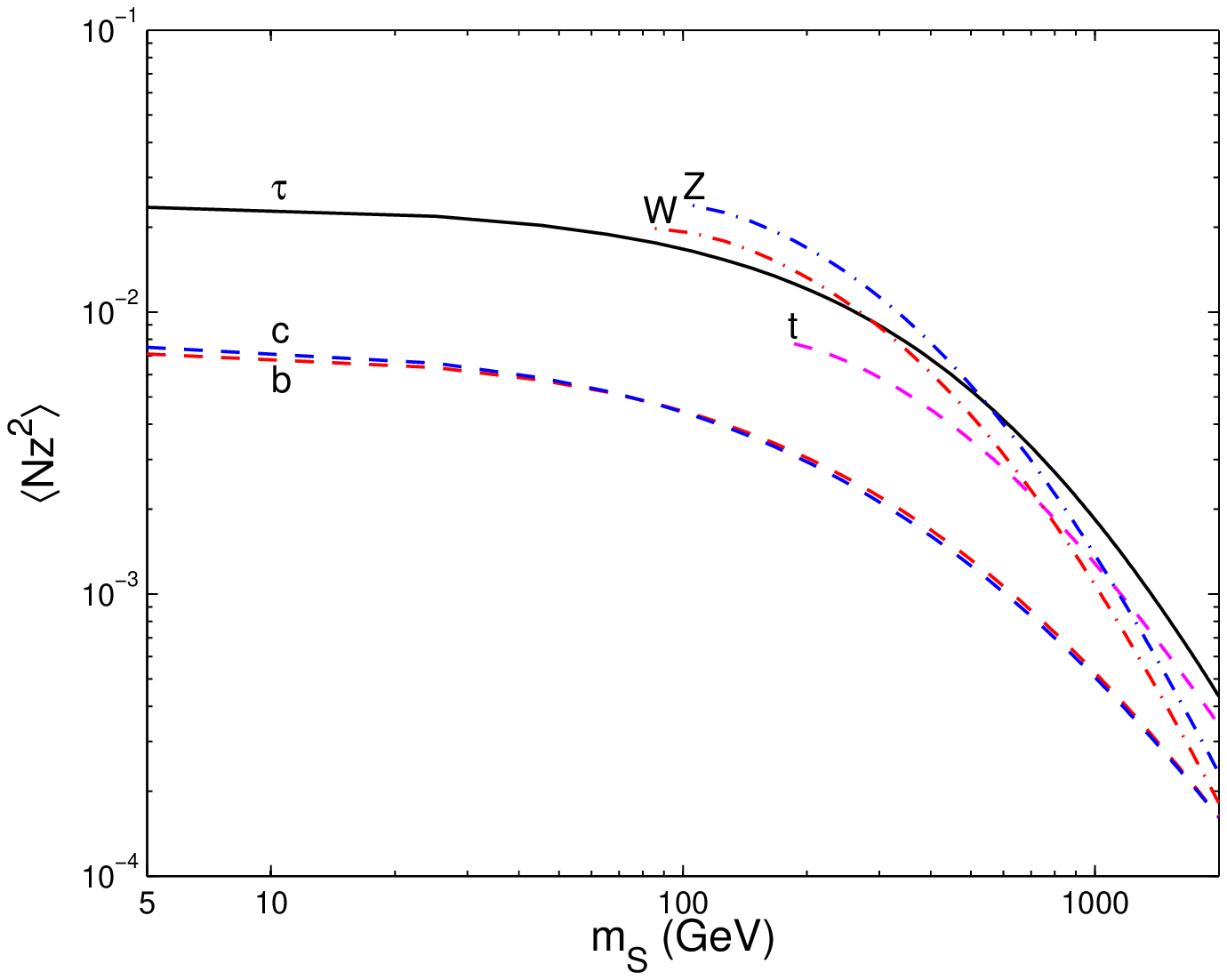}
\includegraphics[width=0.30\textwidth]{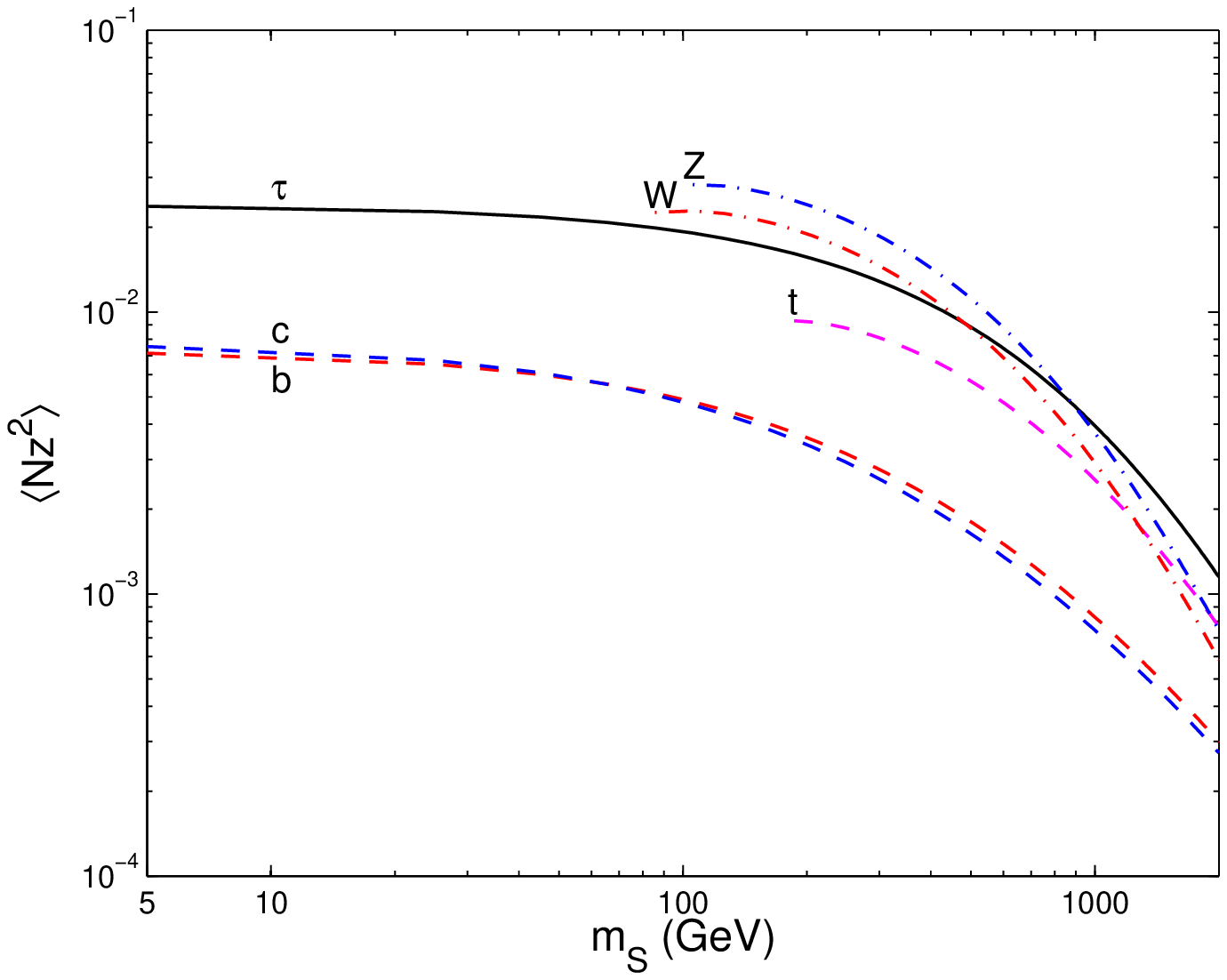} \caption{Second
moment of neutrino spectra from the Earth (left) and the Sun (center).
Also shown is the second moment of anti-neutrino spectrum from the Sun
(right). The Earth anti-neutrino spectrum is identical to the neutrino
spectrum.}\label{fig:nz} \end{figure*}

\subsection{Annihilation} These accumulated DMPs can annihilate into
Standard Model particles. Subsequently, these particles can decay into
energetic muon neutrinos, and hence be detected in astrophysical
neutrino detectors. The scalar $S$ usually annihilates into a two-body
state, and hence the typical energy of the neutrino is $\simeq
\frac{1}{4} m_S$.

The overall annihilation rate in the Sun or Earth is essentially set
by the capture rate (assuming near maximal signal).  We calculate the
dark matter annihilation branching fractions into various Standard
Model particles. We can then use known parametrizations for neutrino
spectra from decay of these particles in the Sun or Earth. The
branching fractions for a specific set of parameters is shown in
Figure \ref{fig:brfr}.

\section{Neutrino Spectra and Detector Rate}

The neutrino spectra from annihilation events depend upon solar
parameters, neutrino physics, quark hadronization
\cite{PhysRevD.34.2206} etc. Once we know the neutrino spectra from
each of the decay products, the overall spectrum is given by
convolving these spectra with the branching fractions. For the
purposes of measurement in the neutrino telescopes, the relevant
quantity is the detector rate. The technique for inferring the
existence of the neutrino is observation of a muon, which is produced
by a charged-current interaction. The cross-section for this process
is proportional to the energy of the neutrino, and the range of the
subsequent muon is also proportional to its energy. Thus, we are
interested in the second moment of the neutrino spectrum for the
detector rate as a function of the injection energy ($E_{in}$)
\cite{PhysRevD.44.3021,Ritz88}. 

\begin{equation} \langle
  Nz^{2}\rangle_{F,i}(E_{in})=\int\left(\frac{dN}{dE}\right)_{F,i}
  (E_{\nu},E_{in})\, \frac{E_{\nu}^{2}}{E_{in}^{2}}\,
  dE_{\nu}\end{equation}

The spectra from the Sun are more complicated than the Earth spectra.
For the Earth, we consider the neutrino spectrum in the rest frame of
the decaying particle, and then boost it for a particle with an energy
$E_{in}$. If the injected particle is a b- or c-quark, we also have to
take hadronization into account. The quark loses energy as it
hadronizes, so the injected energy is a fraction of $E_{in}$.  In the
Solar case, we also have to consider the stopping of heavy hadrons.
The core of the Sun is dense enough to slow b- and c-quarks further
after hadronization. Another effect in the Sun is neutrino stopping
and absorption. Neutrinos lose energy via neutral-current interactions
and can be absorbed through charged-current interactions in the Sun.
Stopping and absorption co-efficients turn out to be different for
neutrinos and anti-neutrinos. Hence, we expect a distinct spectrum for
neutrinos and \mbox{anti-neutrinos} from the Sun. From the Earth,
these spectra are identical. We produce the $\langle Nz^{2}\rangle$
plots in Figure \ref{fig:nz}. Detailed numerical fits for these
functions can be found in \cite{jungman-1996-267}.  The combined
detector rate is given by 

\begin{eqnarray}
  \Gamma_{detect}&=&c\left(\frac{\Gamma_{A}}{\text{s}^{-1}}\right)
  \left(\frac{m_{S}}{\text{GeV}}\right)^{2}\nonumber \\ &\times&
  \sum_{i}a_{i}b_{i}\sum_{F}B_{F}\langle Nz^{2}\rangle_{F,i}(m_{S})
\end{eqnarray}

\begin{figure*}[htbp] \centering
  \includegraphics[width=0.40\textwidth]{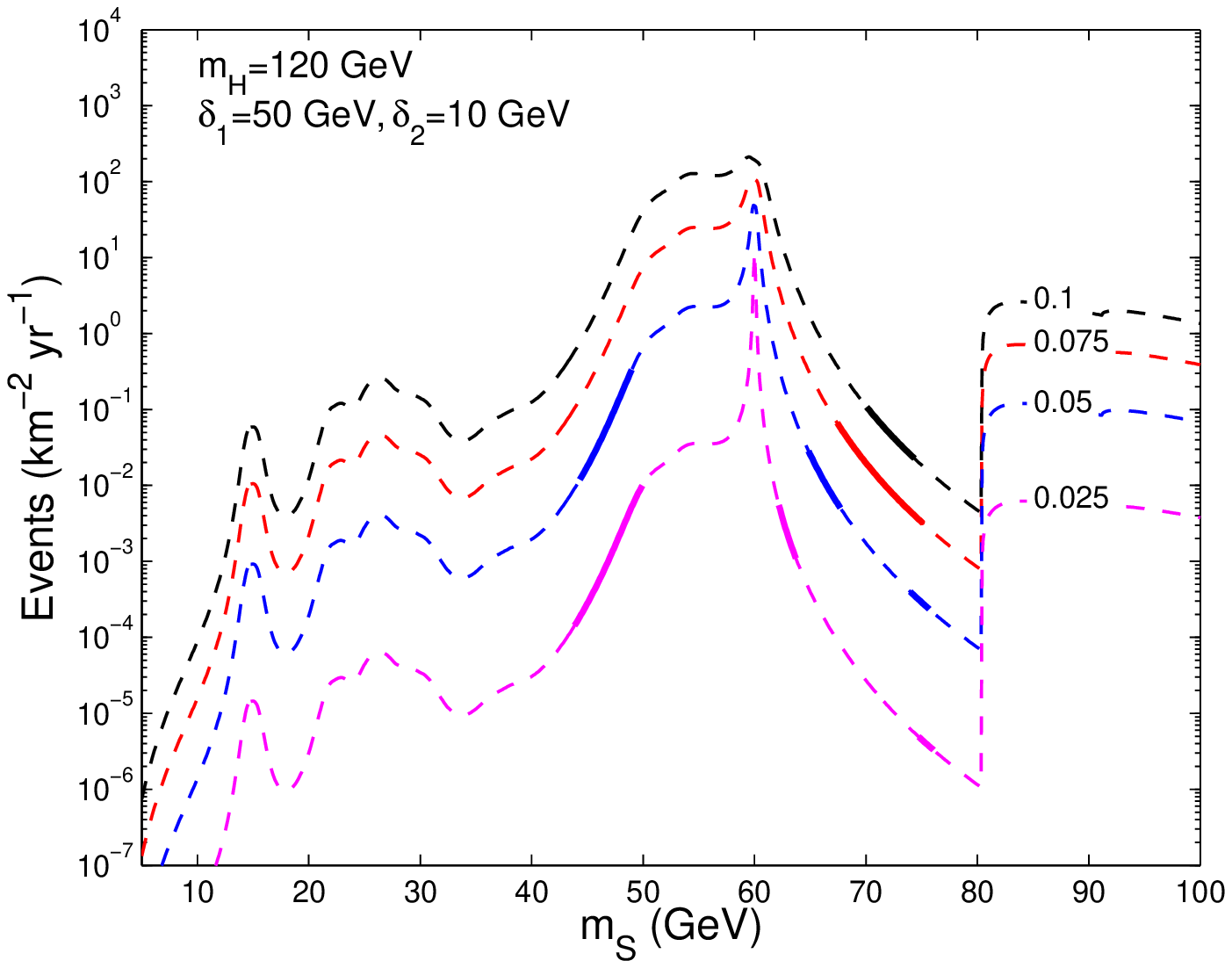}
  \includegraphics[width=0.40\textwidth]{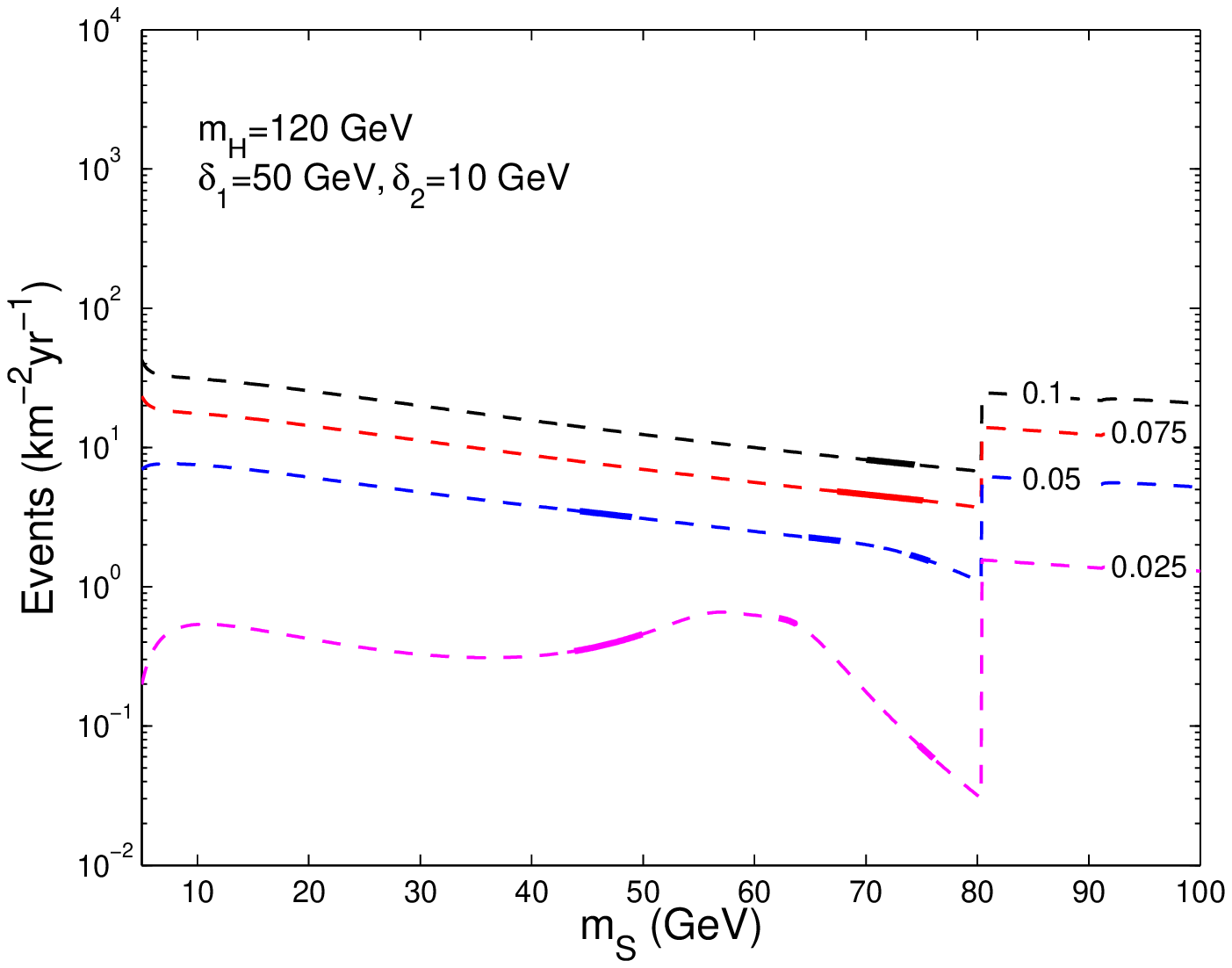}
  \includegraphics[width=0.40\textwidth]{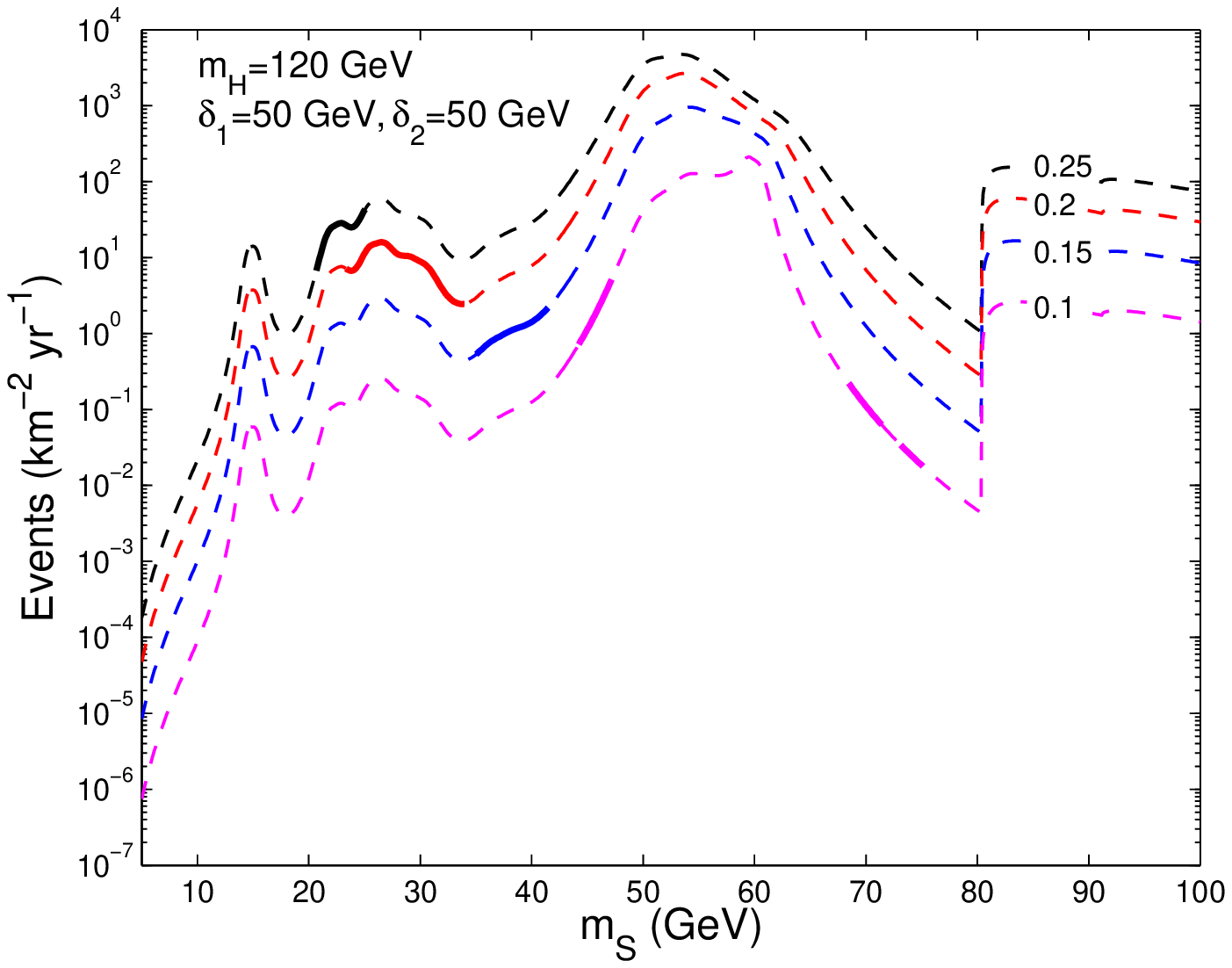}
  \includegraphics[width=0.40\textwidth]{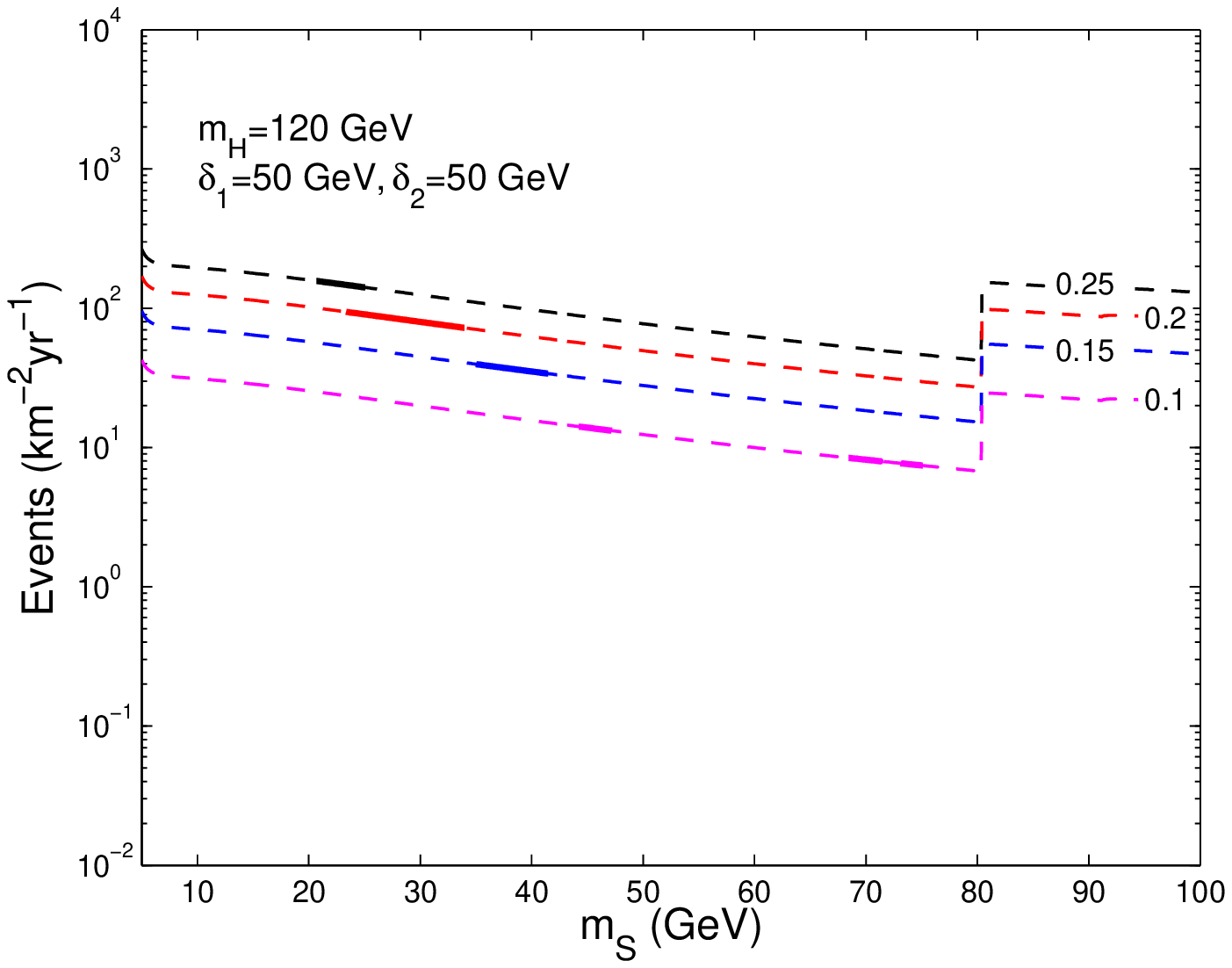}
  \includegraphics[width=0.40\textwidth]{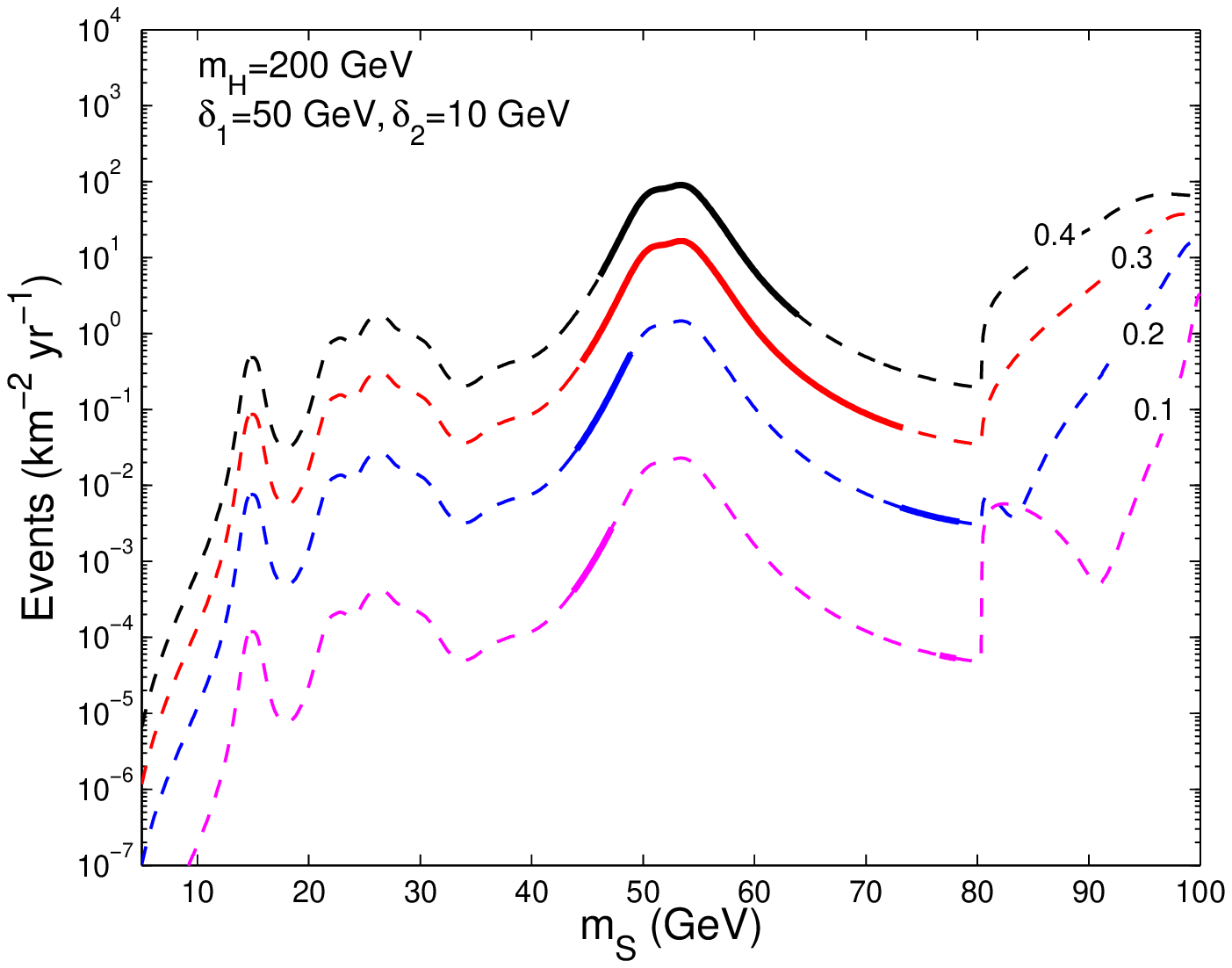}
  \includegraphics[width=0.40\textwidth]{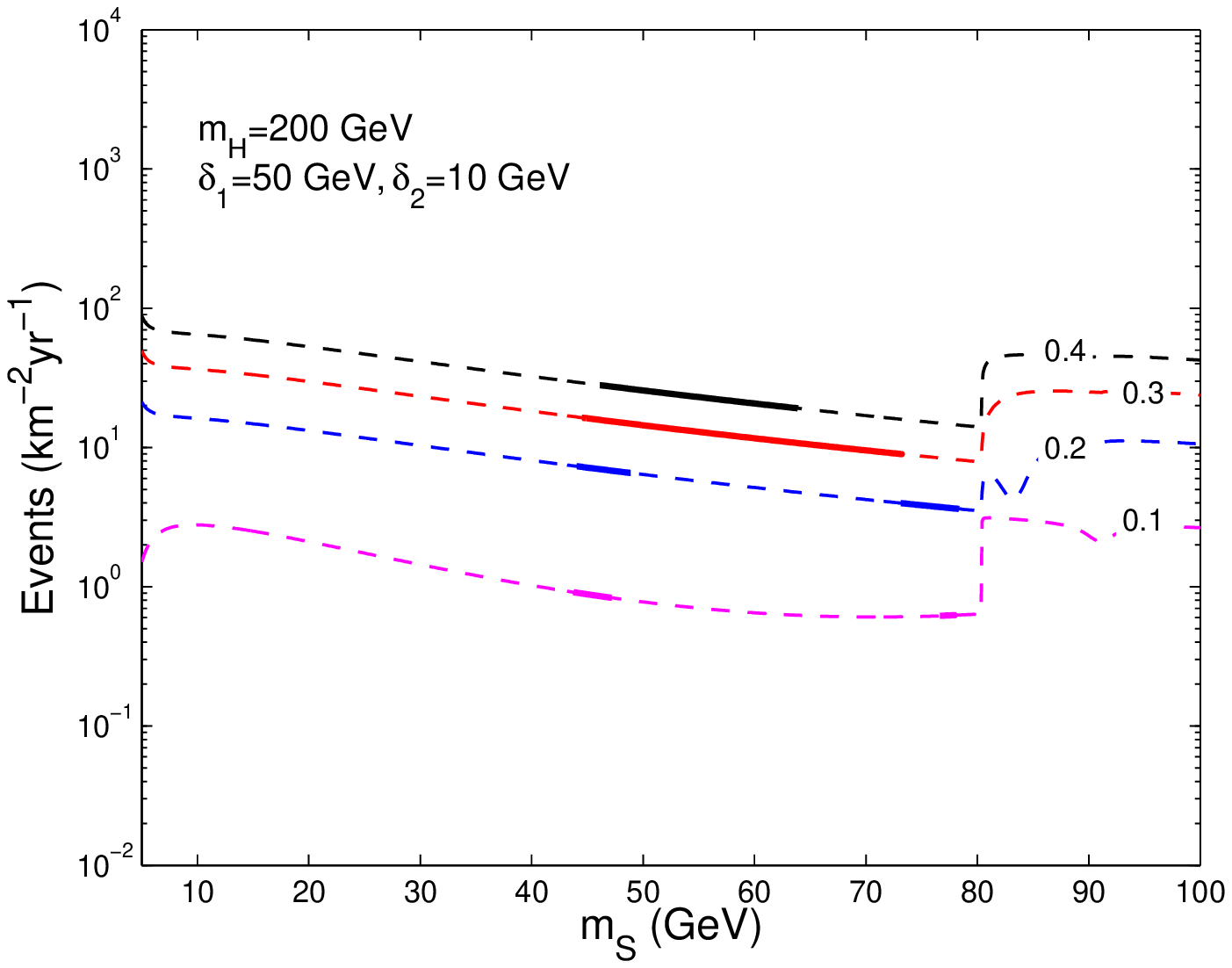}
  \includegraphics[width=0.40\textwidth]{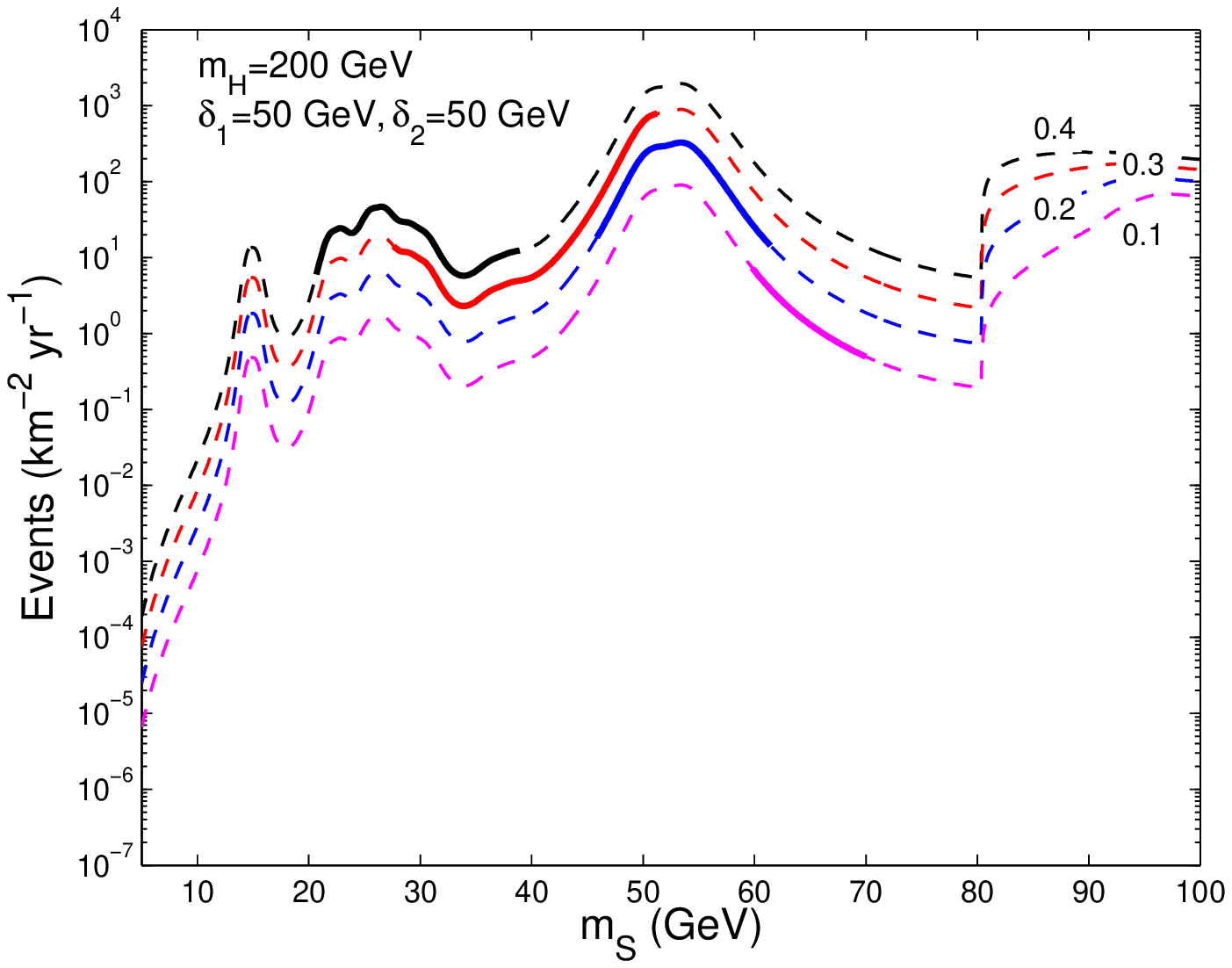}
  \includegraphics[width=0.40\textwidth]{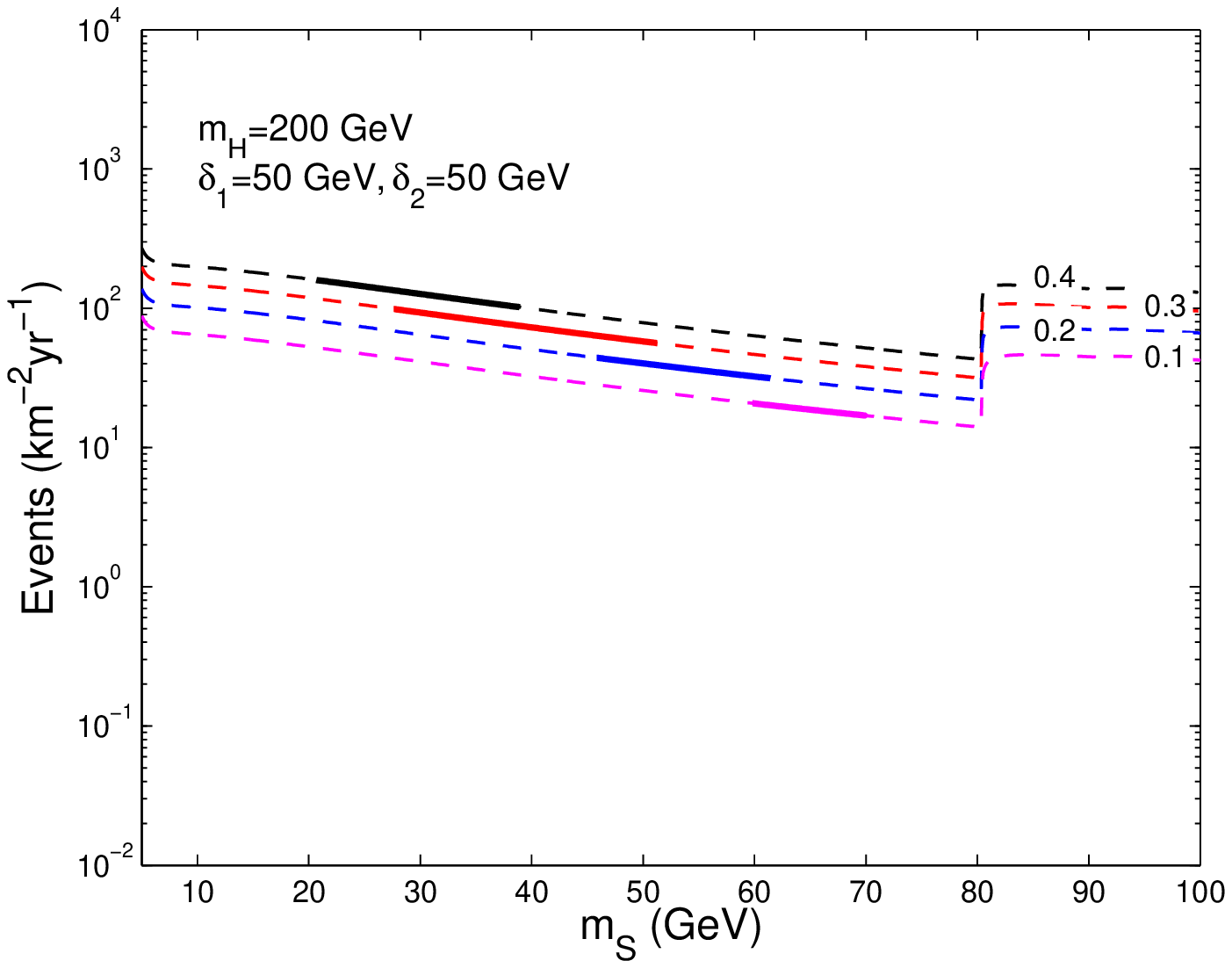}
  \caption{Detector rate from annihilations in the Earth (left) and
  the Sun (right) in the low mass region for various values of
  $\lambda_L$. Different plots are for different Higgs masses and mass
  splittings ($\delta_1$ and $\delta_2$).  The solid sections in the
  lines represent the mass range consistent with the correct relic
  abundance.} \label{fig:drl}

\end{figure*}

\begin{figure*}[htbp] \centering
  \includegraphics[width=0.40\textwidth]{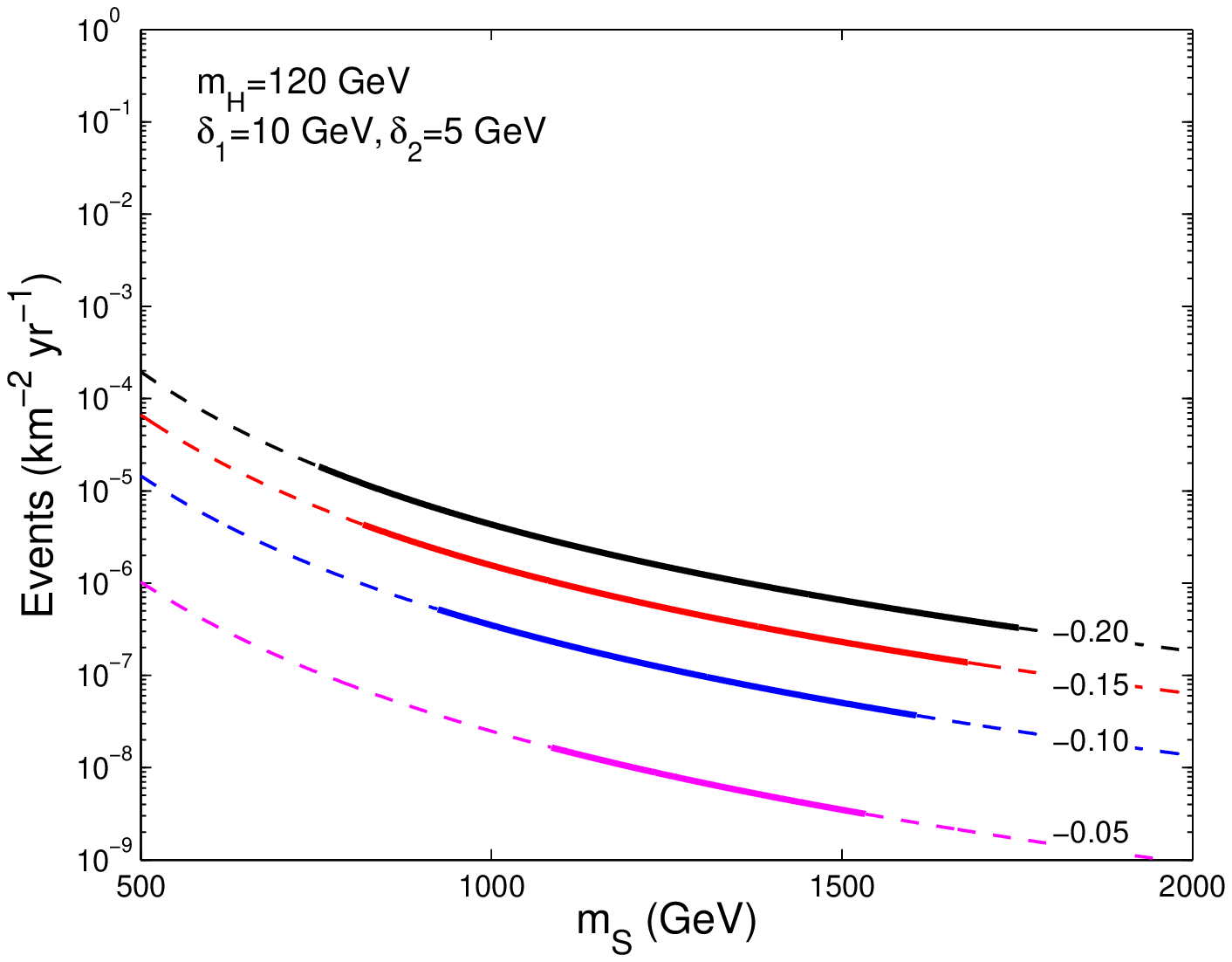}
  \includegraphics[width=0.40\textwidth]{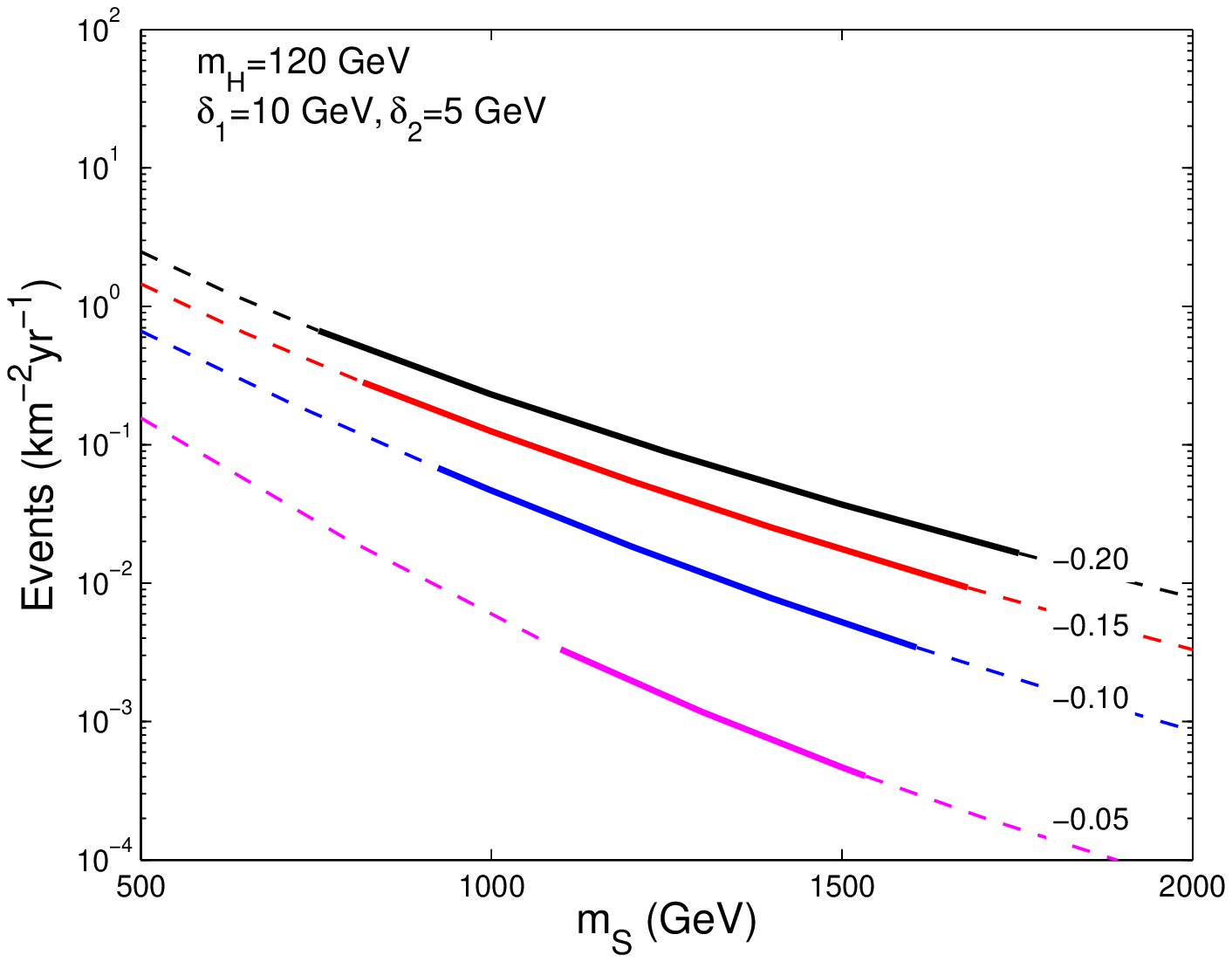}
  \includegraphics[width=0.40\textwidth]{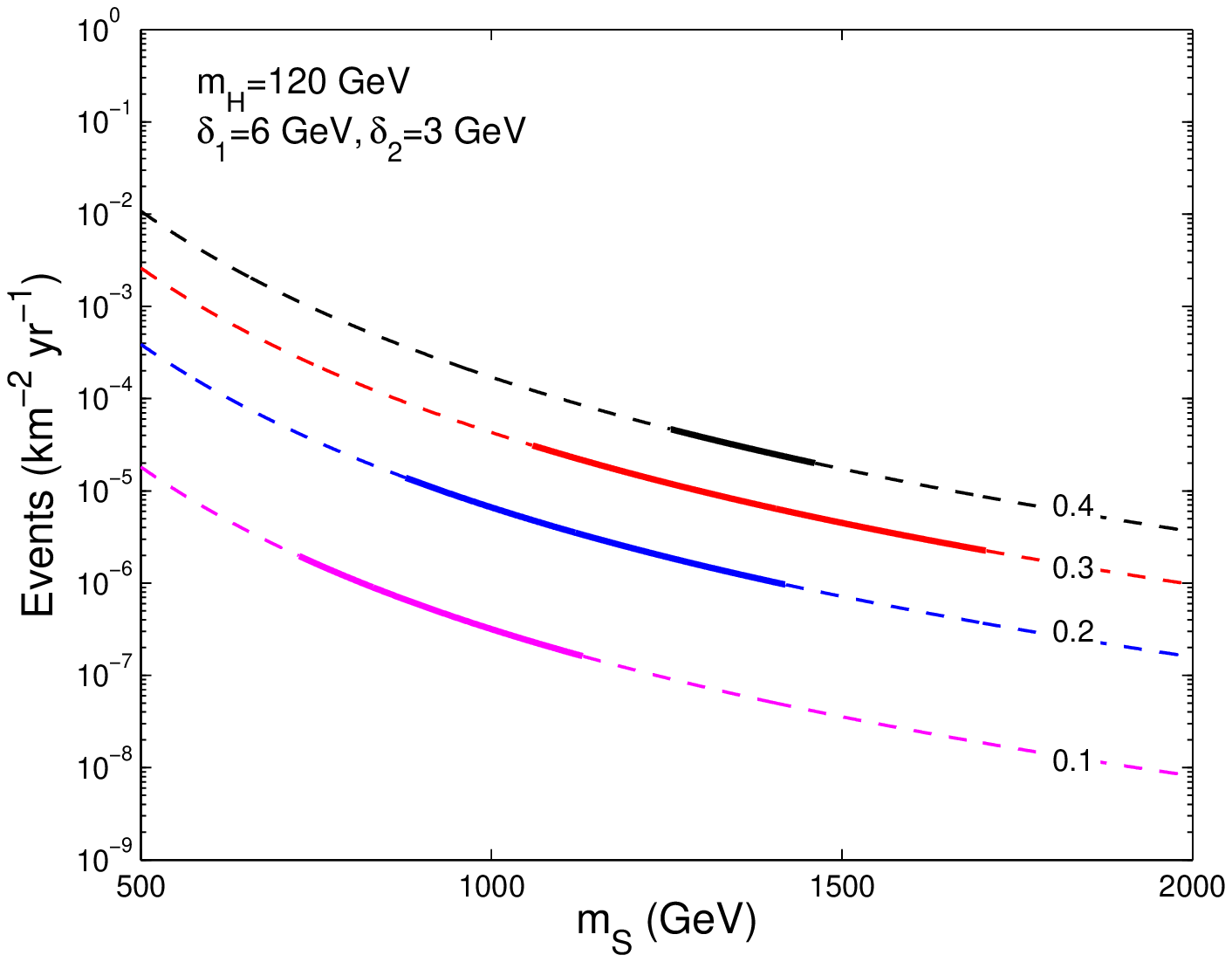}
  \includegraphics[width=0.40\textwidth]{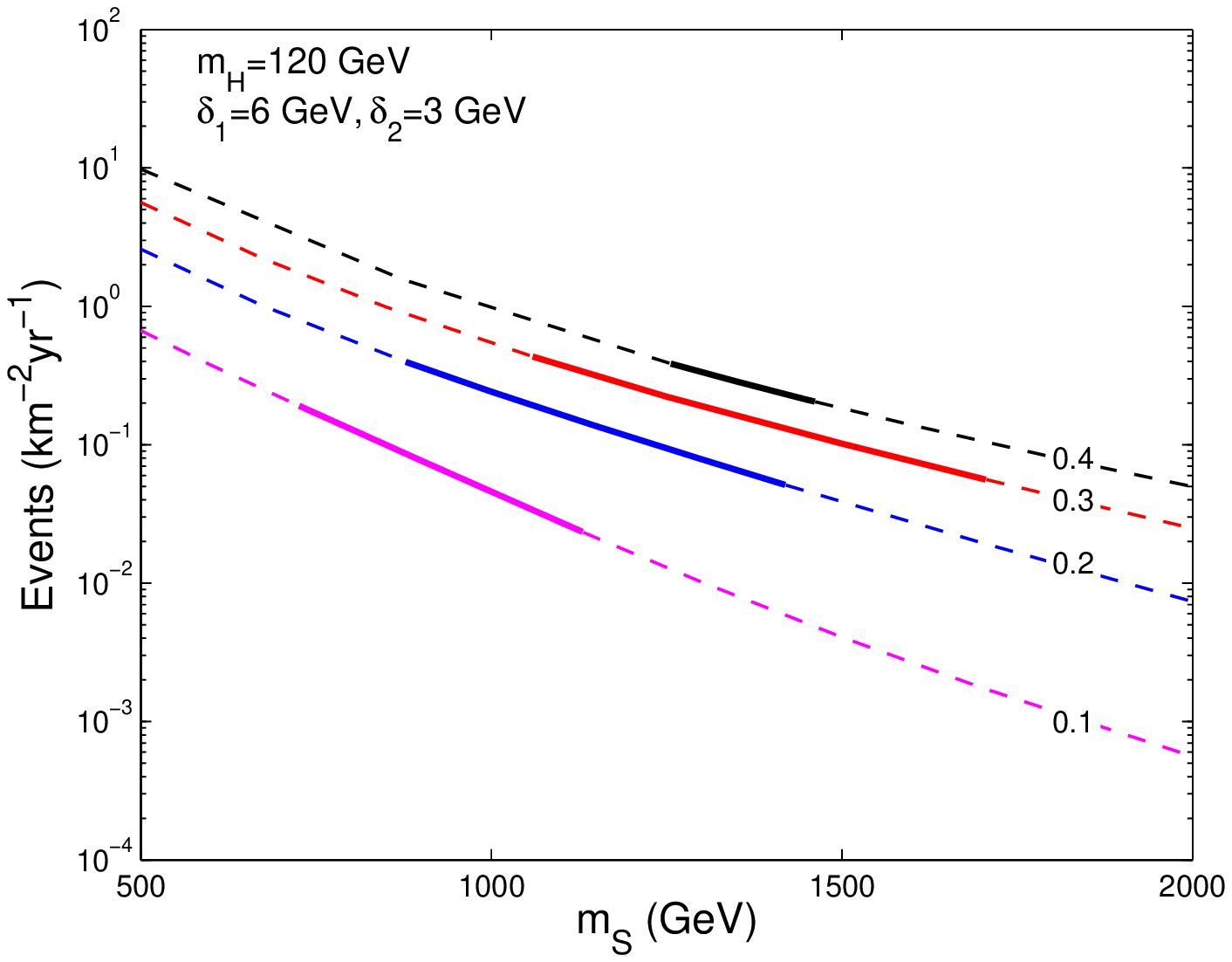}
  \includegraphics[width=0.40\textwidth]{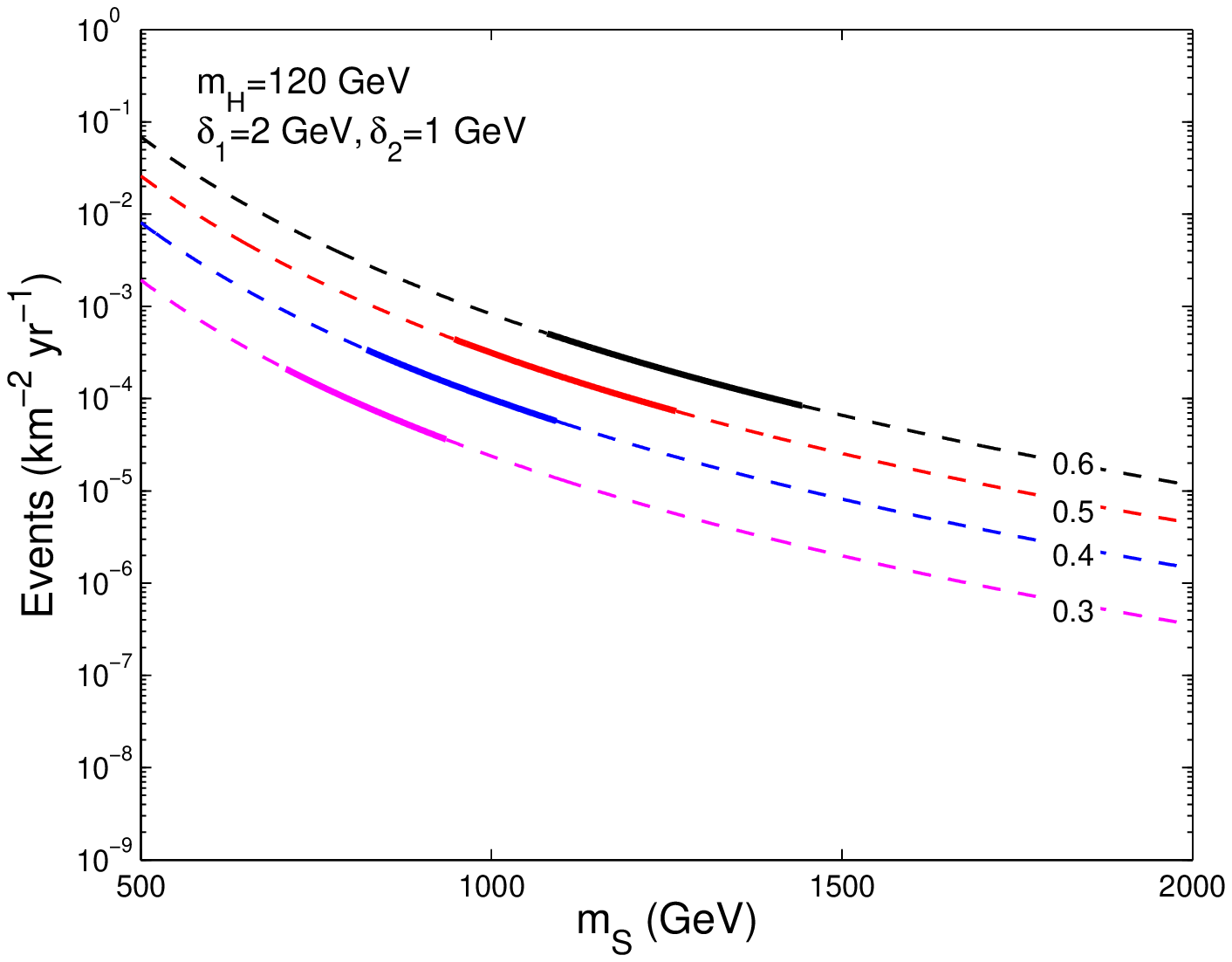}
  \includegraphics[width=0.40\textwidth]{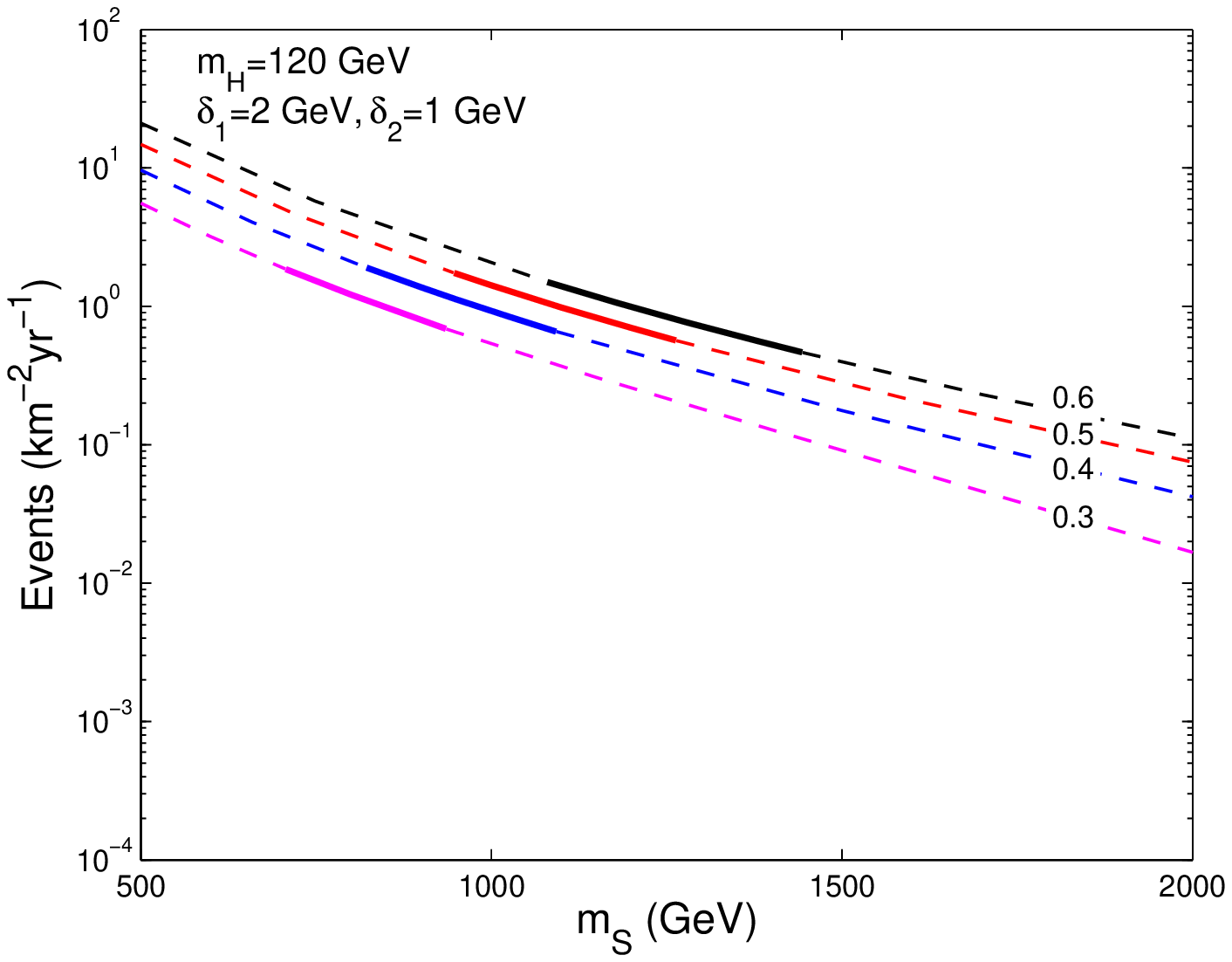}
  \caption{Detector rate from annihilations in the Earth (left) and
  the Sun (right) in the high mass region for various values of
  $\lambda_L$. Different plots are for different mass splittings
  ($\delta_1$ and $\delta_2$).  The solid sections in the lines
  represent the mass range consistent with the correct relic
  abundance.} \label{fig:drh}

\end{figure*}

\afterpage{\clearpage}

The sum over $i$ is over neutrino and anti-neutrino states. The
$a_{i}$ are the neutrino scattering co-efficients, $a_{\nu}=6.8$,
$a_{\bar{\nu}}=3.1$. The $b_{i}$ are the muon range
\mbox{co-efficients}, $b_{\nu}=0.51$ and $b_{\bar{\nu}}=0.67$. $B_{F}$
is the annihilation branching fraction of the DMP for channel $F$.

For the sun, the constant $c = 2.54\times 10^{-23} \text{
km}^{-2}\text{ yr}^{-1}$.  The expression for the detector rate for
the earth is scaled by the square of ratio of the earth-sun distance
to the earth radius. Thus, $c=1.42\times10^{-14}\text{ km}^{-2}\text{
yr}^{-1}$ for the earth.

The background for this process arises mainly from atmospheric
neutrinos \cite{Barger07}.  For the case of the Sun, we can reduce the
background by including events only within a narrow angular cone along
the line of sight from the detector to the Sun. There is also some
background from the solar atmospheric neutrinos \cite{Fogli06}, where
neutrinos are produced in the solar atmosphere by cosmic rays.
\section{Results} We have calculated the expected rate of detection in
neutrino telescopes from these annihilations in the IDM.  We restrict
the parameter space to that allowed by relic abundance constraints by
WMAP and other constraints discussed above. The IDM allows the Higgs
mass to be higher than the standard electro-weak precision test
constraints \cite{Barbieri06}, so we also analyze the signal for a
$200$ GeV SM Higgs.  We do the calculation for different sets of mass
splittings, $\delta_1$ and $\delta_2$.  The plots are produced as a
function of $m_S$ for various values of $\lambda_L$. We find that the
results are qualitatively different for the low mass ($m_S<100$ GeV)
and the high mass ($500\ \text{GeV}<m_S<2\ \text{TeV}$) regions, hence
we present them separately. 

We plot the number of events observed per year in a detector with an
effective area of $1\text{ km}^2$, such as IceCube. The solid sections
on the lines indicate the mass range which gives the correct relic
abundance.

\subsection{Low Mass Region} In the low mass region, the signals from
the Earth and the Sun are comparable. The maximal signal for Earth is
obtained around $m_{S}\simeq50$ GeV. The various peaks in the Earth
plots are explained by kinematics. When the DM mass is close to a
nucleus mass there is no kinematic suppression (Eq.
(\ref{eq:kinsup1})). As the DM mass increases, the cross section for
DM capture decreases, leading to a reduction in the rate. In the Solar
case, there is no kinematic resonance, therefore the Earth signal
exceeds the Solar signal at certain $m_S$ values. 

The branching fractions in the low mass region (below the W
threshold), where annihilation products are quarks ($b$ and $c$) and
leptons ($\tau$), are independent of $\lambda_{L}$. The detector rate
depends on $\lambda_{L}$ only through $\langle\sigma_A v\rangle$ and
the capture rate $(C)$ in this region.

We see $\sim 100$ events from the Earth with a $200$ GeV Higgs and
tens of events with a $120$ GeV Higgs (Figure \ref{fig:drl}).  We
expect to see a few hundred events per year from the Sun in this
parameter range.

\subsection{High Mass Region} In the high mass region, the signal from
Earth is suppressed by several orders of magnitude as compared to the
signal from the Sun. This is because the annihilation rate is not
maximal (the capture and annihilation processes have not come into
equilibrium) for the Earth, but it is for the Sun. The detector rate
decreases with the mass of the dark matter particle, for both Earth
and the Sun, due to the cross-sectional dependence on $m_S$ and
kinematic suppression.  The branching fractions for the high mass
particles have a non-trivial dependence on $\lambda_{L}$.
Consequently, the qualitative features of the detector rate are
influenced by the branching fractions and annihilation rate which, in
turn, depend on the specific value of $\lambda_{L}$. 

In this mass range, the signal from the Earth is too low to be
observed. The signal from the Sun is more promising, and we expect to
see a few events per year.

\section{Conclusion} Neutrino telescopes provide a very interesting
mechanism for dark-matter detection. With the increased sensitivity of
IceCube, these signals can be used to test large ranges of parameter
space in various models. We find a promising signal in the inert
doublet model in two distinct parameter ranges.  We get a few hundred
events per year in the low mass region, and a few events from the Sun
in the high mass region.  For detection, a careful analysis of signal
over background may be required. 

\section{Acknowledgements} We thank Zackaria Chacko and Shufang Su for
valuable discussions.  E.M.D is supported by the DOE under grant
DE-FG02-04ER-41298. P.A.  and C.A.K. are supported by the NSF under
grant PHY-0801323.

\bibliographystyle{apsrevM} 
\bibliography{dark-matter}

\end{document}